\documentstyle[preprint,aps,psfig]{revtex}
\begin{document}
\title{Gravitational Origin of Quark Masses in an Extra-Dimensional
Brane World}

\author{David Dooling and Kyungsik Kang}

\address{Department of Physics, Brown University, Providence RI 02912
USA, and\\
School of Physics, Korea Institute for Advanced Study, Seoul 130-012, Korea
\\ E-mail: dooling@het.brown.edu, kang@het.brown.edu}

\maketitle

\begin{abstract}
Using the warped extra dimension geometry of the many-brane extension of the Randall-Sundrum solution, we find a natural explanation for the observed quark masses of the three Standard Model (SM) generations.
  Localizing massless SM matter generations on neighboring 3-branes in an extra dimensional world leads to phenomenologically acceptable effective four dimensional masses arising from the coupling of the fermion field with the background metric. 
 Thus this geometry can simultaneously address the gauge and quark mass hierarchy problems.
\end{abstract}
\tightenlines
\section{Introduction}

The Standard Model (SM) provides an elegant mechanism by which the massive intermediate vector bosons $W^{\pm}$ and $Z$ acquire mass while the photon and gluons remain massless. 
Postulating the Higgs field to transform as a singlet under $SU(3)_{c}$ and a doublet under $SU(2)$, the $W^{\pm}$ and $Z$ masses at tree level are given in terms of $g_{1}$, $g_{2}$ and only one dimensionful parameter $v \sim 246$ GeV.
The SM does not provide such an economical explanation for the observed fermion masses.   
After spontaneous symmetry breaking, the quark mass term in the lagrangian reads 
\begin{equation}
\mathcal{L_{\mathnormal{mass}}} \mathnormal = \frac{v}{\sqrt{2}} \left( \overline{u_{Li}} h_{ij}^{(u)} u_{Rj} + \overline{d_{Li}} h_{ij}^{(d)} d_{Rj} \right) + h.c.
\label{eq:one}
\end{equation}
where the $h_{ij}$ are arbitrary $3 \times 3$ complex Yukawa coupling matrices.
Some predictive mechanism for fermion mass generation is needed so as to place the understanding of fermion mass on a par with that of gauge boson mass; i.e., to find an overriding principle that predicts the values of the quark and lepton mass matrix elements to be what they are.

Another failing of the SM is that of explaining in a natural way the existence of a light fundamental Higgs mass scale in comparison to the Planck scale, the gauge hierarchy problem.
Traditional proposed solutions have been technicolor and supersymmetry, while recently it has been suggested that large extra dimensions (LED) may result in a change in how gravity behaves at high energies and thus allow room for only one fundamental scale in physics, the TeV scale ~\cite{LED,APL}.
However, this suggestion in its simplest forms does not really address the gauge hierarchy problem, but rather transforms it into a problem of disparate length scales. One needs to explain why the extra dimensions are so large. 
Randall and Sundrum observed that if there are in fact extra dimensions, our world is necessarily confined to a four dimensional submanifold, a 3-brane ~\cite{rs}.
It is then apparent that any 3-branes living in extra dimensions must be taken into account when determining the metric.
The imposition of four-dimensional Poincare invariance generically results in a non-facorizable geometry with an associated warp factor.
Unlike in the LED scenarios, the gauge hierarchy problem is resolved not with the single fundamental scale being the TeV scale, but with the fundamental scale being the Planck scale.
Particle physics scales of 1 TeV are reproduced after taking into account the effect of the warp factor on the visible brane and canonically normalizing the Higgs field, so that a Higgs VEV  $\sim 246$ GeV may result even if the fundamental $V_{0}$ is of the order of the Planck scale. 

In this paper, we ask the question ``can the gauge hierarchy problem and the quark mass hierarachy be simultaneously explained in a minimal extension of the Randall-Sundrum solution?''.
In a recent paper ~\cite{dd}, we used the Lykken-Randall scenario ~\cite{lr} of one hidden, positive tension brane located at the origin in a non-compact extra dimension.
Treating as probes the branes where the SM fields are localized, we found that a phenomenologically acceptable quark mass spectrum results, but that the mixing parameters were inconsistent with experiment.
Taking this discrepency as a signal for a more rigorous treatment, we refine the previous calculation.
Using the full metric as determined by all the branes, including the previously dubbed probe branes, we solve exactly for the fermion field profile in the extra compact dimension.
This new implementation of the same essential mechanism as in \cite{dd} is much more restrictive, and yet we still find the quark mass spectrum can be successfully produced and the gauge hierarchy problem resolved.

The outline of this paper is as follows.
In Section II, we review the many-brane extension of the Randall-Sundrum solution \cite{h} and solve for the profile of a massless fundamental five-dimensional fermion field in this geometry, as outlined in Bajc and Gabadadze \cite{bg}.
In Section III, we present our mechanism of a gravitational origin of quark masses, made possible by the identification of different SM flavors with the peaked profile of one fundmental field around different branes.
This idea of localizing different flavors at different locations in higher-dimensional geography has been exploited recently in \cite{ds,s}.
 We present an example justifying our claim that variants of the RS scenario are capable of successfully addressing both hierarchies mentioned above.
Finally, we generalize the Goldberger-Wise mechanism of modulus stabilization with bulk fields to the multi-brane system under consideration.
We then draw our conclusions and briefly mention further directions currently under investigation.

\section{Many-brane Extension of RS}

In \cite{h}, Hatanaka et al. have generalized the orginal RS solution with an $S^{1}$ compact extra dimension to the case of many branes.
The general configuration considered is that of $N$ parallel 3-branes in five spacetime dimensions such that the $i^{th}$ 3-brane located at $\phi_{i}$ has tension $V_{i}$ $(i=1, 2, ..., N)$ and $0 < \phi_{1} < \phi_{2} < ... < \phi_{N} < 2 \pi$.
This more general configuration allows for the possibility of different inter-brane bulk cosomological constants, and so the entire action may be written as
\begin{equation}
S = S_{grav} + \sum_{i=1}^{N} S_{i}
\end{equation}
where
\begin{equation}
S_{grav} = \int d^{4}x \int_{0}^{2 \pi} \sqrt{G} \left\{ 2 M^{3} R -\sum_{i=1}^{N} \lambda_{i} \left[ \theta \left( \phi - \phi_{i} \right) - \theta \left( \phi - \phi_{i+1} \right) \right] \right\}
\end{equation}
\begin{equation}
S_{i} = \int d^{4}x \sqrt{-g^{(i)}} \left\{ \mathcal{L_{\mathnormal{i}}} \mathnormal - V_{i} \right\}
\end{equation}
and $\theta$ is the Heaviside step function. 
The resulting five dimensional Einstein equations are
\begin{equation}
\sqrt{G} \left( R_{MN} - \frac{1}{2} G_{MN} R \right) = -\frac{1}{4 M^{3}} \left[ \sum_{i=1}^{N} \Lambda_{i} \left[ \theta \left( \phi - \phi_{i} \right) - \theta \left( \phi - \phi_{i+1} \right) \right] \sqrt{G} G_{MN} \right.
\end{equation}
\begin{displaymath}
+ \sum_{i=1}^{N} \left. V_{i} \sqrt{-g^{(i)}} g_{\mu \nu}^{(i)} \delta_{M}^{\mu} \delta_{N}^{\nu} \delta \left( \phi - \phi_{i} \right) \right]
\end{displaymath}
Taking the same form for the metric ans\"{a}tze as in the original RS scenario so as to preserve four dimensional Poincare invariance,
\begin{equation}
ds^{2} = e^{-2 \sigma \left( \phi \right)} \eta_{\mu \nu} dx^{\mu} dx^{\nu} - r_{c}^{2} d \phi^{2}
\end{equation}
one finds the solution
\begin{equation}
\sigma \left( \phi \right) = \left( \lambda_{1} - 0 \right) \left( \phi - \phi_{1} \right) + \left( \lambda_{2} - \lambda_{1} \right) \left( \phi - \phi_{2} \right) \theta \left( \phi - \phi_{2} \right)
\end{equation}
\begin{displaymath}
+ ... + \left( \lambda_{N} - \lambda_{N-1} \right) \left( \phi - \phi_{N} \right) \theta \left( \phi - \phi_{N} \right)
\end{displaymath}
where $S^{1}$ periodicity $( \sigma (0) = \sigma (r_{c} 2 \pi) )$ requires
\begin{equation}
\sum_{i=1}^{N} \lambda_{i} \left( \phi_{i+1} - \phi_{i} \right) = 0
\end{equation}
and $\lambda_{i} = \pm \sqrt{\frac{-\Lambda_{i} r_{c}^{2}}{24 M^{3}}}$, $\frac{V_{i} r_{c}}{12 M^{3}} = \lambda_{i} - \lambda_{i-1}$ for $(i=1, 2, ..., N)$ and $\lambda_{0} = \lambda_{N}$.

Given this general metric, we now wish to couple a fundamental, massless five-dimensional fermion field to it and solve for its profile in the bulk.
In \cite{bg}, Bajc and Gabadadze showed that massless fermions can be localized around a single negative tension brane, but that no normalizable solution exists in the case of a single positive tension brane when the extra dimension is non-compact.
However, in the case of a compact extra dimension, both negative and positive tension branes may exist and the fermion profile in the bulk will have local maxima around the negative tension branes and local minima around the positive tension branes.

Switching notation so that the extra dimension is parametrized by $y$ $\left( 0 < y < y_{c} \right)$, we work with the configuration of twentyfour branes all with the same magnitude of tension, such that
\begin{equation}
\sigma \left( y \right) = \frac{1}{12} \sqrt{ \frac{-6 \Lambda}{M^{3}}} y \theta \left( y \right) + \frac{1}{12 M^{3}} \sum_{i=2}^{24} V_{i} \left( y - y_{i} \right) \theta \left( y - y_{i} \right)
\label{eq:sigma}
\end{equation}

where $V_{i} = \mp V$ as $i$ is even or odd and
\begin{equation}
y_{c} = \frac{V}{ \left( 12 M^{3} \left( \frac{V}{12 M^{3}} - \frac{1}{12} \sqrt{ -\frac{6 \Lambda}{M^{3}}} \right) \right)} \sum_{i=2}^{24} \beta_{i} y_{i}
\end{equation}
where $\beta_{i} = \mp 1$ as $i$ is even or odd.

Given this metric, the properly normalized massless fermion field profile in the bulk is given by
\begin{equation}
\psi = \frac{1}{\sqrt{n}} e^{2 \sigma (y) }
\end{equation}
where $n = \int_{0}^{y_{c}} dy e^{ \sigma (y)} $.
The effective Newton constant is given by
\begin{equation}
M_{pl}^{2} = M^{3} y_{c} \int_{0}^{y_{c}} dy e^{-2 \sigma (y)}
\end{equation}
\begin{equation}
M_{pl}^{2} = M^{3} y_{c} \left[ \left( \frac{1}{2 \lambda_{1}} - \frac{1}{2 \lambda_{N}} \right) + \sum_{i=1}^{N} \left( \frac{1}{2 \lambda_{i}} - \frac{1}{2 \lambda_{i+1}} \right) e^{-2 \sigma (y)} \right]
\end{equation}
and so is of the order $M_{pl}^{-2}$ for every brane.
Hence the $y_{i}$ can be chosen so as to generate an acceptable quark mass spectrum without conflicting with the observed strength of gravity. We note that our multi-brane set-up is a specific example of the so-called brane-crystals as recently investigated by Kaloper \cite{kp}.

\section{Gravitational Origin of Quark Masses}

As in \cite{dd}, we may naturally define an effective four-dimensional pair of quark mass matrices as
\begin{equation}
M_{ij} = \frac{1}{2} \int_{o}^{y_{c}} dy e^{-4 \sigma (y)} \left( \psi_{iL} \left( \partial_{y} \psi_{jR} \right) - \left( \partial_{y} \psi_{iL} \right) \psi_{jR} \right)
\label{eq:mass}
\end{equation}
Up and down quark sector mass matrices may then be evaluated once we adopt a particular brane number - SM field dictionary.
Separating left and right-handed components, we may identify brane number and flavor as $(2,4,6,8,10,12,14,16,18,20,22,24) \rightarrow (d_{R}, u_{R}, d_{L}, u_{L}, s_{R}, c_{R}, s_{L}, c_{L}, b_{R}, t_{R}, b_{L}, t_{L} )$.
Different flavors are identified with different local maxima of the fermion field in the extra dimension.

We now show an example supporting our claim that this geometry allows for a simultaneous taming of the two mass hierarchy problems.
Working in units of the Planck mass, we choose the fundamental parameters as follows: $M=1$, $\Lambda=-1$ and $V=5.51$.
The position of the first brane is at the origin of the compact dimension, $y_{1}=0$.
We then place the remaining branes at equal coordinate intervals further out in the $y$ direction.
In this example, we choose $y_{2} = 49, y_{3} = 56, ..., y_{24} = 203$.
No new hierarchy is introduced, as $y_{c}$ is caculated to be $\sim 227$ Planck lengths.

We show in Fig. (1) a plot of the function $\sigma (y)$ of Eq.~(\ref{eq:sigma}).
The original RS scenario would provide the same picture up to $y_{2} = 49$, the position of the first negative tension brane. 
In this case, the presence of the additional branes of alternating positive and negative tensions of equal magnitude allows for the appearance of the several local maxima and minima of the function $\sigma (y)$.
In Fig. (2) and Fig. (3), we plot the properly normalized fermion field profile in the bulk for this particular choice of parameters.
We see that the peaks associated with the lighter flavors have a markedly higher profile than those associated with the heavier flavors.
Within the context of our proposed mechanism for quark mass generation, this observation makes perfect sense.
Because the fermion field profile goes like the inverse of the warp factor, wherever the fermion profile is large, the interaction with the background metric is dampened, and wherever the fermion profile is small, the interaction with the background metric is enhanced.

%\begin{figure}
%\figurebox{20pc}{15pc}{sigma.ps}
%\epsfxsize=10pc
%\epsfbox{sigma.ps}
%\caption{$\sigma (y)$}
%\end{figure}

\begin{figure}
\psfig{figure=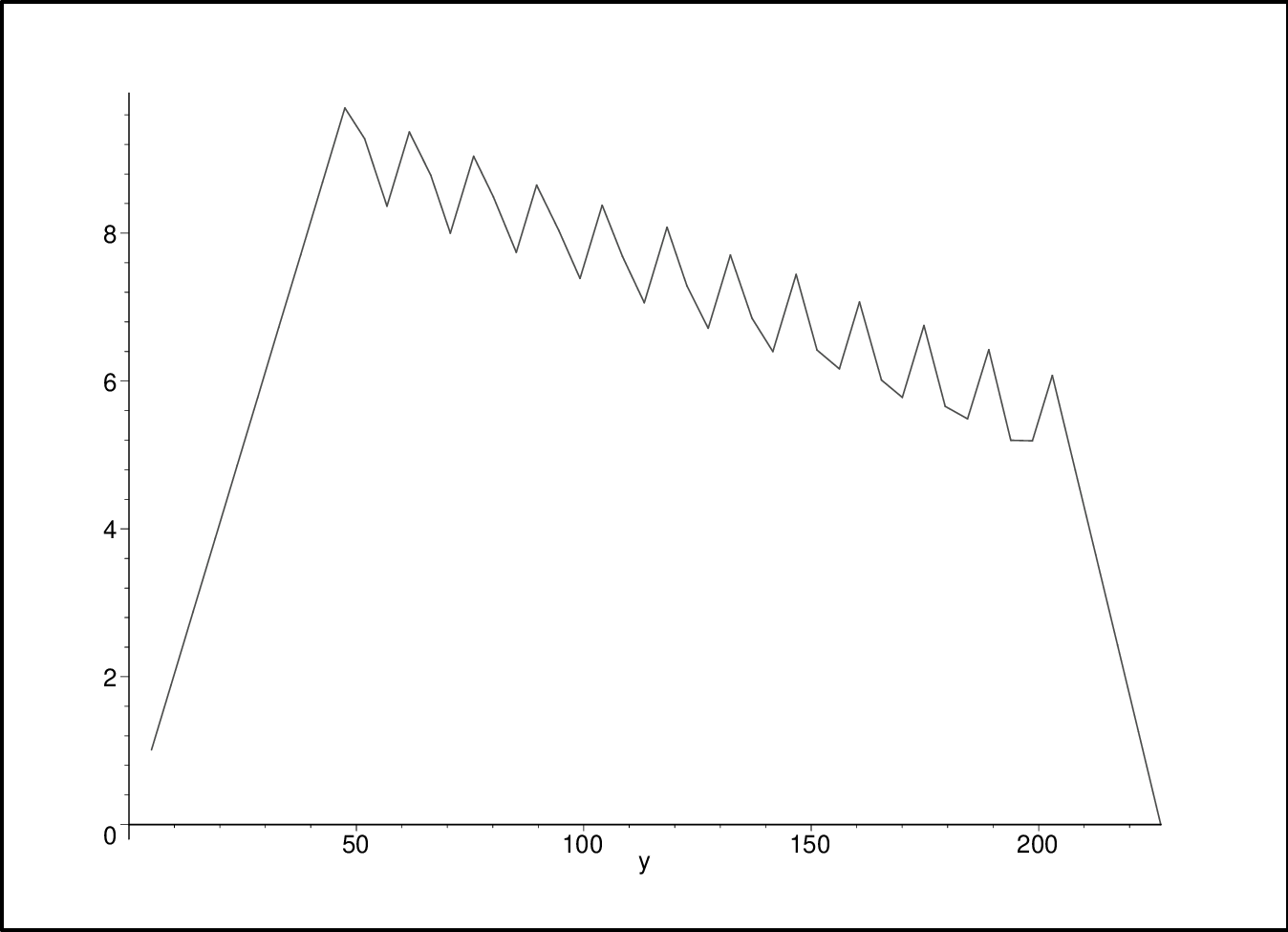,height=9cm,angle=-90}
\begin{quote}
\scriptsize Fig. (1) The function $\sigma (y)$ plotted over the entire $S^{1}$ extra dimension.
\end{quote}
\label{fig:one}
\end{figure}

\begin{figure}
\psfig{figure=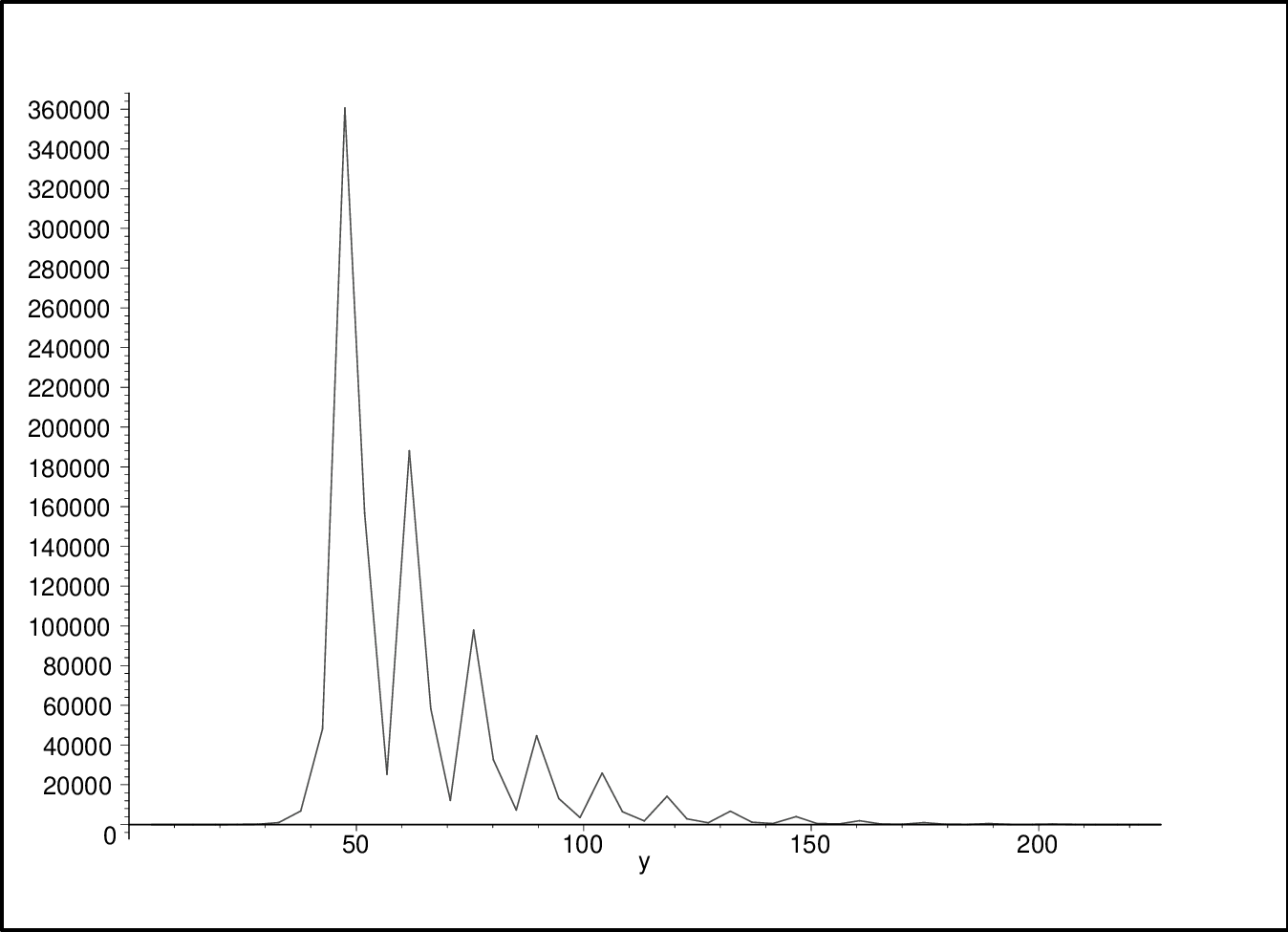,height=9cm,angle=-90}
\begin{quote}
\scriptsize Fig. (2) The properly normalized fermion field profile plotted over the entire $S^{1}$ extra dimension.
\end{quote}
\label{fig:two}
\end{figure}

\begin{figure}
\psfig{figure=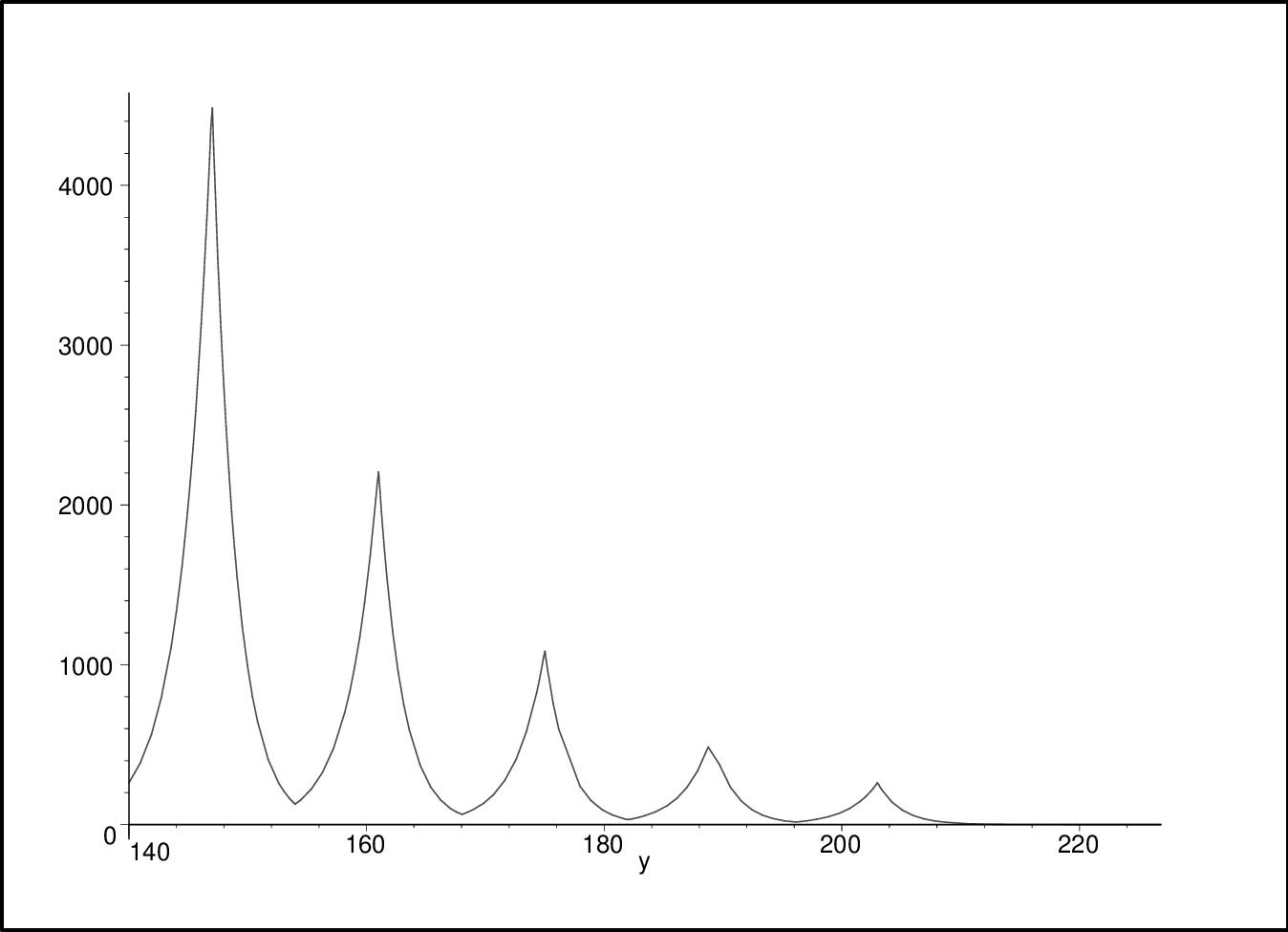,height=9cm,angle=-90}
\begin{quote}
\scriptsize Fig. (3) The properly normalized fermion field profile plotted over the interval of the extra dimension around which the heavier flavors are peaked.
\end{quote}
\label{fig:three}
\end{figure}

Using Eq.~(\ref{eq:mass}) and the above brane number - flavor dictionary, we can compute $\mathcal{M_{\mathnormal u}} \mathnormal = M_{u} M_{u}^{\dagger}$ and $\mathcal{M_{\mathnormal d}} \mathnormal = M_{d} M_{d}^{\dagger}$.

\begin{equation}
\mathcal{M_{\mathnormal u}} \mathnormal = \left( \begin{array}{ccc}
.177432 \times 10^{-40} & .249962 \times 10^{-38} & -.460803 \times 10^{-37} \\
.249962 \times 10^{-38} & .797256 \times 10^{-36} & -.137539 \times 10^{-34} \\
-.460803 \times 10^{-37} & -.137539 \times 10^{-34} & .238164 \times 10^{-33} \end{array} \right)
\label{eq:mup}
\end{equation}
\begin{equation}
 \mathcal{M_{\mathnormal d}} \mathnormal = \left( \begin{array}{ccc}
.102262 \times 10^{-41} & .144233 \times 10^{-39} & -.265891 \times 10^{-38} \\
.144233 \times 10^{-39} & .460510 \times 10^{-37} & -.794454 \times 10^{-36} \\
-.265891 \times 10^{-38} & -.794454 \times 10^{-36} & .137567 \times 10^{-34} \end{array} \right)
\label{eq:md}
\end{equation}
where we remind the reader that we have been working in units of the Planck mass $M_{pl} \sim 1.221047 \times 10^{19}$ GeV.
Computing the eigenvalues, one finds $m_{t} \sim 188$ GeV, $m_{c} \sim .66$ GeV, $m_{u} \sim .002$ GeV, $m_{b} \sim 45 $ GeV, $m_{s} \sim  .159 $ GeV and $m_{d} \sim .0005$ GeV.
With the exception of the bottom quark, these values are in good agreement with observation and prompt us to take this model seriously. 
The mixing matrix calculated in the usual way from $V_{CKM} = U_{u}^{\dagger} U_{d}$ where $U_{u}^{\dagger} \mathcal{M_{\mathnormal u}} \mathnormal U_{u} = \mbox{diag} \left( m_{u}^{2}, m_{c}^{2}, m_{t}^{2} \right)$ and $U_{d}^{\dagger} \mathcal{M_{\mathnormal d}} \mathnormal U_{d} = \mbox{diag} \left( m_{d}^{2}, m_{s}^{2}, m_{b}^{2} \right)$ is computed to be essentially the $3 \times 3$ identity matrix.
This result is in contrast to our previous implementation of this mechanism in \cite{dd}, in which there was too much mixing between the second and third generations.
In our model, flavor mixing does not arise from the diagonalization of four dimensional quark mass matrices, but rather from wave function overlap, as in \cite{ds}.

As one may readily verify, varying the $y_{i}$ and the other fundamental parameters by factors of the order of 2, or even relaxing the aestheticly pleasing requirements of equal magnitudes for the brane tensions and equal coordinate spacings for the brane positions in the extra dimension, one can indeed successfully produce a quark mass spectrum in all but perfect agreement with experimental observation.
This variation may require some degree of moderate fine-tuning, but nothing like the fine-tuning associated with the gauge hierarchy problem within the context of the pure SM.

\section{Stabilization of the Multi-Brane System}

Because the positions of the various branes are of crucial importance to the above analysis, it is necessary to investigate how the branes can be stabilized at these positions in such a way that no new hierarchy is introduced.
In the orginal RS scenario, $r_{c}$ is taken to be the vacuum expectation value of a scalar field.
Because this scalar field did not have an associated potential, it was not clear that the size of the extra dimensions would be stable.
Goldberger and Wise (GW) \cite{gw} then solved this problem by introducing a massive bulk scalar field $\Phi$, the radion, with quartic self-interactions localized on the hidden and visible 3-branes.
The coupling of $\Phi$ on the branes causes $\Phi$ to develop an expectation value that varies with position in the extra dimension.
Finding a solution for $\Phi$ and then integrating the action for $\Phi$ with this solution over the extra dimension yields an effective four-dimensional potential for $r_{c}$.
In the limit of vanishing radion bulk mass and assuming that $\Phi$ minimizes the potential on the branes, this effective four-dimensional potential has a minimum at
\begin{equation}
k r_{c} = \left( \frac{4}{\pi} \right) \frac{k^{2}}{m^{2}} \ln \left[ \frac{v_{h}}{v_{v}} \right] .
\end{equation}
Thus, the introduction of the radion field can stabilize the distance between the two branes in a way that does not introduce additional hierarchies.

Choudhury et al. \cite{mah} have generalized the GW mechanism to the three 3-brane system of Kogan et al. \cite{kog}
In this section, we show how the same mechanism can stabilize the positions of an arbitrary number of 3-branes in an extra compact dimension.

The general configuration considered is that of $N$ parallel 3-branes in five spacetime dimensions such that the $i^{th}$ 3-brane located at $\phi_{i}$ has tension $V_{i} (i=1,2,...N)$ and $0 = \phi_{1} < \phi_{2} < \phi_{3} < ... < \phi_{N} < 2 \pi$.
We restrict ourselves to the case where the branes have alternating positive and negative tensions of equal magnitude, $V_{brane}$, with the brane located at the origin of the extra dimension having positive tension.
The $S^{1}$ symmetry requires that $\sigma (0) = \sigma (2 \pi)$, and thus the solution Eq.(7) implies that $N$ must be even.

In the orginal RS scenario $\sigma ( \phi ) = k r_{c} | \phi |$.
In our multi-brane system in the $S^{1}$ extra dimension, we have different $k_{i}'s$ in the regions between the $i^{th}$ and $i+1^{th}$ branes: $k_{1} = \sqrt{ -\frac{ \Lambda}{24 M^{3}}}, k_{2} = -\sqrt{-\frac{\Lambda}{24 M^{3}}}, k_{3} = \sqrt{-\frac{\Lambda}{24 M^{3}}}, ..., k_{N} = -\sqrt{-\frac{\Lambda}{24 M^{3}}}$ where $N$ is even and our choices of $k_{i}$ demand that all the bulk cosmological constants in between the branes be equal.

To stabilize the $r_{c} \phi_{i}$, we add to the model a scalar field $\Phi$ with the following bulk action:
\begin{equation}
S_{b} = \frac{1}{2} \int \mbox{d}^{4}x \int_{0}^{2 \pi} \mbox{d} \phi \sqrt{G} \left( G^{AB} \partial_{A} \Phi \partial_{B} \Phi - m^{2} \Phi^{2} \right)
\end{equation}
where $G_{AB}$ with $A, B = \mu, \phi$ is given by Eq. (6) with $\sigma(\phi) = \pm kr_{c} \left( \phi\right)$ for $ \phi_{i} < \phi < \phi_{i+1}$ and where $\phi_{i+1} = 2 \pi$.
We also include interaction terms on the branes located at $\phi_{1} = 0, \phi_{2}, ..., \phi_{N}$ given by
\begin{eqnarray*}
S_{1} & = & -\int \mbox{d}^{4} x \sqrt{-g^{(1)}} \lambda_{1} \left( \Phi^{2} - v_{1}^{2} \right)^{2} \\
S_{2} & = & -\int \mbox{d}^{4} x \sqrt{-g^{(2)}} \lambda_{2} \left( \Phi^{2} -v_{2}^{2} \right)^{2} \\
 . & & \\
 . & & \\
S_{N} & = & -\int \mbox{d}^{4} x \sqrt{-g^{(N)}} \lambda_{N} \left( \Phi^{2} - v_{N}^{2} \right)^{2} \\
\end{eqnarray*} 
where $g^{(i)}$ is the determinant of the induced metric on the $i^{th}$ brane.
 The interaction terms on the brane cause the field $\Phi$ to develop a $\phi$-dependent vacuum expectation value which is determined classically by solving the differential equation \cite{gw}
\begin{equation}
0 = -\frac{1}{r_{c}^{2}} \partial_{\phi} \left( e^{-4 \sigma} \partial_{\phi} \Phi \right) + m^{2} e^{-4 \sigma} \Phi + \sum_{i=1}^{N} \frac{4}{r_{c}} e^{-4 \sigma} \lambda_{i} \Phi \left( \Phi^{2} - v_{i}^{2} \right) \delta \left( \phi - \phi_{i} \right).
\end{equation}
In the $i^{th}$ region defined by $ \phi_{i} < \phi < \phi_{i+1}$ between the $i^{th}$ and $i+1^{th}$ branes, the solution is given by
\begin{equation}
\Phi (\phi) = e^{2 \sigma} \left[ a_{i} e^{\nu \sigma} + b_{i} e^{-\nu \sigma} \right]
\end{equation}
with $\nu = \sqrt{ 4 + \frac{m^{2}}{k^{2}}}$ and $\sigma = \pm k r_{c} \phi$ with the positive sign corresponding to odd $i$ and the negative sign to even $i$.

In principle, the coefficients $a_{i}$ and $b_{i}$ are determined by integrating the above equation of motion over  vanishingly small regions centered about the $i^{th}$ and $i+1^{th}$ branes.
If we do not make any assumptions about the values of $\Phi (\phi_{i})$, this method of matching at the $i^{th}$ brane provides only $N + 1$ constraints (N branes on which to match and the condition $\Phi(0) = \Phi( 2 \pi)$ ) for the $2N$ unknowns $a_{i}$ and $b_{i}$.
This counting shows that for $N > 1$, the system is underdetermined.
We therefore make the physically reasonable assumption that $\Phi ( \phi_{i}) = v_{i}$; i.e. that $\Phi$ minimizes the boundary potential.
With this assumption, we find the following expressions for $a_{i}$ and $b_{i}$.
For odd $i$:
\begin{equation}
a_{i} = \frac{ -v_{i} e^{-2 \nu k r_{c} \phi_{i+1}} e^{-2 k r_{c} \phi_{i}} e^{ \nu k r_{c} \phi_{i}} + v_{i+1} e^{-2 k r_{c} \phi_{i+1}} e^{-\nu k r_{c} \phi_{i+1}}}{\left( 1 - e^{-2 \nu k r_{c} (\phi_{i+1} - \phi_{i})} \right) }
\end{equation}
\begin{equation}
b_{i} = \frac{ v_{i} e^{\nu k r_{c} \phi_{i}} e^{-2 k r_{c} \phi_{i}} - v_{i+1} e^{2 \nu k r_{c} \phi_{i}} e^{-\nu k r_{c} \phi_{i+1}} e^{-2 k r_{c} \phi_{i+1}}}{ \left( 1 - e^{-2 \nu k r_{c} ( \phi_{i+1} - \phi_{i})} \right) }
\end{equation}
For even $i$, we find the following expressions for $a_{i}$ and $b_{i}$:
\begin{equation}
a_{i} = \frac{ v_{i} e^{(2 + \nu) k r_{c} \phi_{i}} - v_{i+1} e^{2 \nu k r_{c} \phi_{i}} e^{2 k r_{c} \phi_{i+1}} e^{-\nu k r_{c} \phi_{i+1}}}{ \left( 1 - e^{-2 \nu k r_{c} (\phi_{i+1} - \phi_{i})} \right) }
\end{equation}
\begin{equation}
b_{i} = \frac{ -v_{i} e^{-2 \nu k r_{c} \phi_{i+1}} e^{2 k r_{c} \phi_{i}} e^{ \nu k r_{c} \phi_{i}} + v_{i+1} e^{2 k r_{c} \phi_{i+1}} e^{ -\nu k r_{c} \phi_{i+1}}}{ \left( 1 - e^{-2 \nu k r_{c} (\phi_{i+1} - \phi_{i})} \right) }
\end{equation}
and where $\phi_{N+1} = 2 \pi$.

Putting this solution for $\Phi$ into $S_{b}$ and integrating over $\phi$ yields an effective four-dimensional potential for the $\phi_{i}$ which takes the form $\sum_{i=1}^{N} V_{i}$ where
\begin{eqnarray}
V_{i} & = &\frac{a_{i}^{2}}{2} \left[ \left( e^{2 \nu k_{i} r_{c} \phi_{i+1}} - e^{-2 \nu k_{i} r_{c} \phi_{i}} \right) \left( \nu + 2 \right) k_{i} \right]\\
 & + & \frac{b_{i}^{2}}{2} \left[ \left( e^{-2 \nu k_{i} r_{c} \phi_{i}} - e^{-2 \nu k_{i} r_{c} \phi_{i+1}} \right) \left( \nu - 2 \right) k_{i} \right]
\end{eqnarray}
with $k_{i} = k$ for odd $i$ and $k_{i} = -k$ for even $i$.

 We now suppose that the radion mass is small in the sense that $ \nu = \sqrt{ 4 + \frac{m^{2}}{k^{2}}}$ may be well approximated by $ \nu = 2 + \epsilon$ where $\epsilon \sim \frac{m^{2}}{4k^{2}}$ is small \cite{gw}.
In this limit of small $\epsilon$, we see from the above expression for $V_{i}$ that the term multiplying $b_{i}^{2}$ is already linear in $\epsilon$.
So if we wish to examine the potential $V$ in the vanishing $\epsilon$ limit, it suffices to consider only the first term in the $V_{i}.$ 

To this end, we find the following expressions for the $a_{i}$ where subleading powers of $e^{-kr_{c}( \phi_{i+1} - \phi_{i})}$ have been neglected:
\begin{equation}
a_{i} = e^{-(4+ \epsilon)k r_{c} \phi_{i+1}} \left( v_{i+1} - v_{i} e^{-\epsilon k r_{c} ( \phi_{i+1} - \phi_{i} )} \right)
\end{equation}
\begin{equation}
a_{i} = e^{(4 + \epsilon) k r_{c} \phi_{i}} \left( v_{i} - v_{i+1} e^{ -\epsilon k r_{c} ( \phi_{i+1} - \phi_{i})} \right)
\end{equation}
where the first expression holds for odd $i$ and the second for even $i$.
With these expressions for the $a_{i}$, the potentials $V_{i}$ become
\begin{equation}
V_{i} = 2k \left( 1 + \frac{ \epsilon}{4} \right) e^{-4 k r_{c} \phi_{i+1}} \left( v_{i+1} - v_{i} e^{-\epsilon k r_{c} (\phi_{i+1} - \phi_{i})} \right)^{2} \left( 1 - e^{-(4 + 2 \epsilon) k r_{c} (\phi_{i+1} + \phi_{i})} \right)
\end{equation}
\begin{equation}
V_{i} = 2k \left( 1 + \frac{ \epsilon}{4} \right) e^{12kr_{c} \phi_{i}} e^{ 4 \epsilon k r_{c} \phi_{i}} \left( v_{i} - v_{i+1} e^{-\epsilon k r_{c} ( \phi_{i+1} - \phi_{i} ) } \right)^{2} \left( 1 - e^{-(4 + 2 \epsilon) k r_{c} ( \phi_{i+1} + \phi_{i} ) } \right)
\end{equation}
for odd and even $i$, respectively.

These potentials are minimized when the squared terms vanish, as the total factors multiplying them are positive definite.
It follows that
\begin{equation}
r_{c} \left( \phi_{i+1} - \phi_{i} \right) = \frac{4k}{m^{2}} \ln \left( \frac{v_{i+1}}{v_{i}} \right)
\end{equation}
\begin{equation}
r_{c} \left( \phi_{i+1} - \phi_{i} \right) = \frac{4k}{m^{2}} \ln \left( \frac{v_{i}}{v_{i+1}} \right)
\end{equation}
for even and odd $i$, respectively.
Given the $N$ $v_{i}$ $(v_{N+1} = v_{1})$, the above relations and the $S^{1}$ symmetry requirements that $\phi_{1} = 0$ and $\phi_{N+1} = 2 \pi$, the remaining $\phi_{i}$ are then stabilized.
Using the parameters of the example given in Section 3, where $y_{i} = r_{c} \phi_{i}$, one may readily check that no new hierarchies among the $v_{i}$ are introduced for small $\epsilon$.
That this basic generalization of the GW mechanism may be applied to our model is not surprising; the only extra twist is the necessity of treating even and odd $i$ seperately as a result of having opposite sign slopes for the funcion $\sigma (\phi)$ in adjacent regions.

\section{Conclusions and Further Directions}

We have found a consistent, straightforward mechanism within the extra dimensional scenario to derive effective four dimensional quark mass matrices that are phenomenologically acceptable.
Several issues warrant further investigation to either lend more credence to this model or to ban it to the bonfires of happy mathematical coincidences.
The first issue is that of the stability of the brane coordinates, which enter strongly into the determination of the mass eigenvalues.
Presumably, the simple generalization of the Goldberger-Wise mechanism could account for their stability.
A more sophisticated explanation would not only account for their stability, but for the actual values needed to produce an acceptable mass spectrum.
An additional point that needs to be addressed is our model's lack of CP violation.
As is known, the SM mechanism for CP violation via a physically meaningful phase in the CKM matrix does not provide enough CP violation for the purposes of baryogenesis.
Perhaps the source of CP violation and the origin of quark masses are different problems.
To conclude, the RS scenario provides an intriguing resolution of the hierarchy problem.
We find suggestive evidence that it may address the fermion mass hierarchy as well.
Mass being the charge of spacetime-matter interactions, it seems only natural that the fermion mass hierarchy will be understood in terms of spacetime considerations as opposed to an internal flavor symmetry governing Yukawa-type interactions.

\section*{Acknowledgements}
We wish to thank the Korea Institute for Advanced Study for warm hospitality during the completion of this work.
DD also gratefully acknowledges the U.S. Dept. of Ed. for financial support via the Graduate Assistance in Areas of National Need (GAANN) program and the NSF for support via a Dissertation Enhancement Award (INT-0083352).
Support for this work was also provided in part by the U.S. Dept. of Energy grant DE-FG02-91ER40688. Institutional report numbers for this work are BROWN-HET-1239, BROWN-TA-586 and KIAS-P00069.

\end{document}